\documentstyle[11pt,epsfig]{article}
\begin{document}
\begin{center}
\boldmath
{\Large\bf Semileptonic decays $B\to(\pi,\rho)e\nu$ in relativistic quark model}
\\
\vspace{.5cm}
{\large D.~Melikhov}
\\ 
\vspace{.5cm}
{\normalsize {\it Nuclear Physics Institute, Moscow State University, Moscow,
119899, Russia} \\ Electronic address: melikhov@monet.npi.msu.su}
\end{center}
\unboldmath

Quark model results for the $B\to \pi,\rho$ decays are analysed, 
making use of the dispersion formulation of the model:  
The form factors at $q^2>0$ are expressed as 
relativistic invariant double spectral representation over 
invariant masses of the initial and final mesons through their light--cone wave functions. 
The dependence of the results on the quark model parameters is studied. 
For various versions of the quark model the ranges  
$\Gamma(\bar B^0\to\pi^+e\bar\nu)=(7\pm2)\times10^{12}|V_{ub}|^2\; s^{-1}$, 
$\Gamma(\bar B^0\to\rho^+e\bar\nu)/\Gamma(\bar B^0\to\pi^+e\bar\nu)=1.45\pm0.1$, and 
$\Gamma_L/\Gamma_T=0.7\pm0.08$ are found. The effects of the constituent quark transition 
form factor are briefly discussed. \\
PACS number(s): 13.20.-v, 12.39.Hg, 12.39.Ki.

\vspace{.5cm}

Weak decays of hadrons provide an important source of information on 
the parameters of the standard model of electroweak interactions, the structure of
weak currents,  
and internal structure of hadrons. Hadron decay rates involve both the 
Cabibbo--Kobayashi--Maskawa matrix elements and hadron form factors, 
therefore the extraction of the standard model parameters from the experiments
on hadron decays requires reliable information on hadron structure. 

A theoretical study of hadronic matrix elements of the weak currents inevitably 
encounters the problem of describing the hadron structure and requires 
a nonperturbative consideration. This gives the main uncertainty to the theoretical
predictions for hadron transition amplitudes. 

For the transitions between hadrons each containing a heavy quark and light degreess of freedom, 
the number of the independent form factors is considerably reduced due to the heavy quark
symmetry \cite{iw}. For instance, in the leading $1/m_Q$ order the transition between heavy mesons  
is described in terms of the single Isgur--Wise function, which should be estimated within a  
nonperturbative approach. The $O(1/m^N_Q)$ corrections to this picture can be
consistently calculated within the Heavy Quark Effective Theory \cite{hqet}, 
an effective theory based on QCD in the limit of large quark masses 
(a detailed review can be found in \cite{neubert}).
  
For the decays caused by the heavy--to--light quark transitions the 
argumentation of the heavy--to--heavy transitions does not work, and the 
situation turns out to be  much less definite. 
For instance, for the decays $B\to \pi,\rho$ the uncertainty of the results of nonperturbative 
approaches such as the quark model \cite{wsb}--\cite{faustov}, QCD sum rules \cite{sr1}--\cite{sr3}, 
and lattice calculations \cite{lat1}--\cite{lat3} is too large to draw any definite 
conclusion on the form factor values and decay rates 
(see Tables \ref{table:b2piffs}, \ref{table:b2pirates}). 

A step forward in the understanding of heavy--to--light transitions was recently done by B.Stech
who noticed that relations between heavy--to--light form factors can be 
obtained if use is made of the constituent quark picture \cite{stech}. Namely, assuming  
that (i) the momentum distribution of the constituent quarks inside a meson is strongly peaked with a
width corresponding to the confinement scale, and (ii) the process in which the spectator 
retains its spin and momentum dominates the transition amplitude, 
\footnote{ Actually, one more assumption on the dynamics of the procees is employed. Namely, the
picture in \cite{stech} includes both the constituent and current quarks. And for deriving the final
relations it is important that the momentum of the current quark coincides with the momentum of the 
corresponding constituent quark. This assumption allows one to avoid the appearence of the constituent
quark transition form factor which should be taken into account if the picture with only 
constituent quarks is considered.}
the 6 form factors can be reduced to a single function just 
as it is in the case of the heavy--to--heavy transition. These relations are expected to be valid 
up to the corrections $O\left(2m_{u}M_B/(M_B^2+M_\pi^2-q^2)\right)$. Although these corrections cannot be 
estimated numerically, the relations can be a guideline to the 
analysis of the heavy--to--light decay processes. 
  
Constituent quark picture has been extensively applied to the description of the decay processes 
\cite{wsb}--\cite{faustov}, \cite{jaus}. 
Although in the first models by 
Wirbel, Stech, and Bauer (WSB) \cite{wsb} and Isgur, Scora, Grinstein and Wise (ISGW) 
\cite{isgw} quark spins were not treated 
relativistically, it has become clear soon that 
for a consistent application of quark models to electroweak decays,  
a relativistic treatment of quark spins is necessary \cite{neubert}. 
The exact solution to this complicated dynamical problem is not known, but a simplified
self--consistent relativistic treatment of the quark spins can be performed within 
the light--cone formalism \cite{lcqm}. 
The only difficulty with this approach is that the applicability of the model is 
restricted by the condition $q^2\le 0$, while 
the physical region for hadron decays is 
$0\le q^2\le (M_i-M_f)^2$, $M_{i,f}$ being the initial and final hadron mass, respectively. 
So, for obtaining the form factors in the physical region and decay rates and lepton
distributions, assumptions on the form factor
$q^2$--behavior were necessary. 
A procedure to remedy this difficulty has been proposed in \cite{m}. 

The approach of \cite {m} is based on the dispersion formulation of the light--cone quark model 
\cite{amn}. 
Namely, the transition form factors obtained within the light--cone formalism at $q^2<0$ 
\cite{jaus}, are represented as dispersion 
integrals over initial and final hadron masses. The transition form factors at $q^2>0$ are 
derived by performing the analytic
continuation in $q^2$ from the region $q^2\le 0$. As a result, for a decay caused by the weak 
transition of the quark $Q(m_i)\to Q(m_f)$, 
form factors in the region $q^2\le (m_i-m_f)^2$ are expressed through the light--cone 
wave functions of the initial and final hadrons. 
We apply this approach to the analysis of the $B\to(\pi,\rho)$ decays   
and study the dependence of the
results on the quark model parameters, such as constituent quark masses and wave functions. 
We also check the fulfillment of the Stech relations in particular model calculations.  

The amplitudes of the semileptonic decays of a pseudoscalar meson $P(M_1)$ into the final pseudoscalar
$P(M_2)$ and vector $V(M_2)$ mesons have the following structure \cite{iw}
\begin{eqnarray}
\label{amplitudes}
<P(M_2,p_2)|V_\mu(0)|P(M_1,p_1)>&=&f_+(q^2)(p_1+p_2)_\mu+f_-(q^2)(p_1-p_2)_\mu  \nonumber \\
<V(M_2,p_2,\epsilon)|V_\mu(0)|P(M_1,p_1)>&=&2g(q^2)\epsilon_{\mu\nu\alpha\beta}\epsilon^{*\nu}(p_2)p_1^\alpha
p_2^\beta  \\
<V(M_2,p_2,\epsilon)|A_\mu(0)|P(M_1,p_1)>&=&if(q^2)\epsilon^*_\mu+ia_+(q^2)(\epsilon^*p_1)(p_1+p_2)_\mu+
ia_-(q^2)(\epsilon^*p_1)(p_1-p_2)_\mu   \nonumber 
\end{eqnarray}
We denote both the pseudoscalar and vector meson masses as $M_2$ but the relevant value
is taken in each case. 

Our goal is to calculate the form factors of the transition between $S$-wave mesons at  
$0\le q^2\le(M_1-M_2)^2$ within the constituent quark picture. In this picture the initial meson 
$P(M_1)$ is a bound state of the constituent quarks $Q(m_2)\bar Q(m_3)$, the final meson 
is an $S$-wave bound state $Q(m_1)\bar Q(m_3)$, and the transition process is described by the graph
of Fig.1. 

We however start with the region $q^2<0$ and make use of the fact that 
the transition form factors calculated within the light--cone quark model 
\cite{jaus} can be written as double spectral representations \cite{m},\cite{amn} 
over the invariant masses of the initial and final mesons 
\begin{equation}
\label{ff1}
f_i(q^2)=f_{21}(q^2)
\int\limits^\infty_{(m_1+m_3)^2}\frac{ds_2G_{2}(s_2)}{\pi(s_2-M_2^2)}
\int\limits^{s_1^+(s_2,q^2)}_{s_1^-(s_2,q^2)}\frac{ds_1G_{1}(s_1)}{\pi(s_1-M_1^2)}
\frac{\tilde f_i(s_1,s_2,q^2)}{16\lambda^{1/2}(s_1,s_2,q^2)},
\end{equation}
where 
$$
s_1^\pm(s_2,q^2)=-\frac1{2m_1^2}
$$
$$
\times\left({
s_2(q^2-m_1^2-m_2^2)-q^2(m_1^2+m_3^2)+
(m_1^2-m_2^2)(m_1^2-m_3^2)\pm\lambda^{1/2}(s_2,m_3^2,m_1^2)\lambda^{1/2}(q^2,m_1^2,m_2^2)
}\right)
$$
and
$$
\lambda(s_1,s_2,s_3)=(s_1+s_2-s_3)^2-4s_1s_2.
$$
Here $G_{1,2}$ are the vertex functions which describe the constituent quark structure of the mesons,
$f_{21}(q^2)$ is the form factor of the constituent quark weak transition $m_2\to m_1$. 
In what follows we adopt the conventional approximation $f_{21}(q^2)=1$. 
The same representations of the form factors has been obtained in \cite{akms} 
taking into account the contribution of two--particle singularities in the 
Feynman graphs.  

The double spectral densities $\tilde f_i$ of the form factors $f_+,g,a_+$, and $f$, which give a 
nonvanishing contribution to the 
cross--section of the semileptonic decay in the case of zero lepton mass, 
read (hereafter $s_3\equiv q^2$)
\begin{eqnarray}
\tilde f_+(s_1,s_2,s_3)&=&D+(\alpha_1+\alpha_2)D_3,  \\
\tilde g(s_1,s_2,s_3) &=& 2\left[ m_1\alpha_2+m_2\alpha_1+m_3(1-\alpha_1-\alpha_2)-
\frac{2\beta}{\sqrt{s_2}+m_1+m_3}\right],   \\
\tilde a_+(s_1,s_2,s_3)&=&2\left[{
2m_2\alpha_{11}+2m_3(\alpha_1-\alpha_{11})-m_1\alpha_2-m_2(\alpha_1-2\alpha_{12})}\right.  \\
&&\left.{-m_3(1-\alpha_1-\alpha_2-2\alpha_{12})
+\frac{C\alpha_1+C_3(\alpha_{11}+\alpha_{12})}{\sqrt{s_2}+m_1+m_3}}\right], \nonumber  \\
\tilde f(s_1,s_2,s_3)&=&\frac{M_2}{\sqrt{s_2}}\tilde f_{D}(s_1,s_2,s_3)+
M_2\tilde a_+(s_1,s_2,s_3)
\left({\frac{s_1-s_2-s_3}{2\sqrt{s_2}}-\frac{M_1^2-M_2^2-s_3}{2M_2}}\right), 
\end{eqnarray}
$$
\tilde f_D(s_1,s_2,s_3)=4\left[E+2\beta(m_2-m_3)+\frac{C_3\beta}{\sqrt{s_2}+m_1+m_3}\right]
$$
where
$$
D=s_1+s_2-(m_2-m_3)^2-(m_1-m_3)^2,\quad D_3=s_3-(m_1-m_2)^2-D,
$$
$$
C=s_1+s_2-(m_2-m_3)^2-(m_1+m_3)^2,\quad C_3=s_3-(m_1+m_2)^2-C,
$$
$$
E=m_1m_2m_3+\frac{m_2}2(s_2-m_1^2-m_3^2)
+\frac{m_1}2(s_1-m_2^2-m_3^2)-\frac{m_3}2(s_3-m_1^2-m_2^2),
$$
$$
\alpha_1=\frac1{\lambda(s_1,s_2,s_3)}
\left[(s_1+s_2-s_3)(s_2-m_1^2+m_3^2)-2s_2(s_1-m_2^2+m_3^2)\right],
$$
$$
\alpha_2=\frac1{\lambda(s_1,s_2,s_3)}
\left[(s_1+s_2-s_3)(s_1-m_2^2+m_3^2)-2s_1(s_2-m_1^2+m_3^2)\right],
$$
$$
\beta=\frac14\left[2m_3^2-\alpha_1(s_1-m_2^2+m_3^2)-\alpha_2(s_2-m_1^2+m_3^2)\right],
$$
$$
\alpha_{11}=\alpha_1^2+4\beta\frac{s_2}{\lambda(s_1,s_2,s_3)}, \quad
\alpha_{12}=\alpha_1\alpha_2-2\beta\frac{s_1+s_2-s_3}{\lambda(s_1,s_2,s_3)}.
$$
For an $S$-wave meson (vector and pseudoscalar) with the mass $M$ built up of the 
constituent quarks $m_1$ and $m_2$, the function $G$ is normalized as follows \cite{m}
\begin{equation}
\label{vertnorm}
\int\frac{G^2(s)ds}{\pi(s-M^2)^2}\frac{\lambda^{1/2}(s,m_1^2,m_2^2)}{8\pi s}(s-(m_1-m_2)^2)=1. 
\end{equation}

Notice that the double dispersion representation without subtractions are valid
for the form factors $f_+,\;f_-\;g,\;a_+$, and $a_-$, while the form factor $f$ requires
subtractions. 
%\footnote{This is a general situation for any amplitude with a vector particle in the
%initial or final state. Such an amplitude is transverse with respect to the momentum of the 
%vector particle, and this yields a necessity of subtractions in some of the form factors.}

Now, we obtain the form factors at $q^2>0$ by performing the analytic continuation in $q^2$.
For the function which has at $q^2\equiv s_3\le 0$ the structure 
\begin{equation}
%\label{ff2}
\phi(s_3)=
\int\limits^\infty_{(m_1+m_3)^2}\frac{ds_2G_{2}(s_2)}{\pi(s_2-M_2^2)}
\int\limits^{s_1^+}_{s_1^-}\frac{ds_1G_{1}(s_1)}{\pi(s_1-M_1^2)}
\left[\frac{P_0(s_1,s_2,s_3)}{\lambda^{1/2}(s_1,s_2,s_3)}
+\frac{P_1(s_1,s_2,s_3)}{\lambda^{3/2}(s_1,s_2,s_3)}
+\frac{P_2(s_1,s_2,s_3)}{\lambda^{5/2}(s_1,s_2,s_3)}\right],
\end{equation}
where $P_i$ are polynomials of $s$, the analytical continuation to the region $q^2>0$ yields
the following expression at $q^2\le(m_2-m_1)^2$
\begin{equation}
\label{ff2}
\phi(s_3)=
\int\limits^\infty_{(m_1+m_3)^2}\frac{ds_2G_{2}(s_2)}{\pi(s_2-M_2^2)}
\int\limits^{s_1^+}_{s_1^-}\frac{ds_1G_{1}(s_1)}{\pi(s_1-M_1^2)}
\left[\frac{P_0(s_1,s_2,s_3)}{\lambda^{1/2}(s_1,s_2,s_3)}
+\frac{P_1(s_1,s_2,s_3)}{\lambda^{3/2}(s_1,s_2,s_3)}
+\frac{P_2(s_1,s_2,s_3)}{\lambda^{5/2}(s_1,s_2,s_3)}\right]
\end{equation}
$$
+2\theta(s_3)
\int\limits^\infty_{s_2^0}\frac{ds_2G_{2}(s_2)}{\pi(s_2-M_2^2)}
\int\limits^{\infty}_{s_1^R}\frac{ds_1}{\pi(s_1-s_1^R)^{1/2}}
\left[\tilde\phi_0(s_1)+\frac{\tilde\phi_1(s_1)-\tilde\phi_1(s_1^R)}{s_1-s_1^R}
+\frac{\tilde\phi_2(s_1)-\tilde\phi_2(s_1^R)-\tilde\phi'_2(s_1^R)(s_1-s_1^R)}
{(s_1-s_1^R)^2}\right],
$$
\begin{equation}
\sqrt{s_2^0}=-\frac{s_3+m_1^2-m_2^2}{2\sqrt{s_3}}+
\sqrt{
\left({  
\frac{s_3+m_1^2-m_2^2}{2\sqrt{s_3}}
}\right)^2+(m_3^2-m_1^2) 
},\quad s_3<(m_2-m_1)^2,
\end{equation}
$s_1^L=(\sqrt{s_2}-\sqrt{s_3})^2$, $s_1^R=(\sqrt{s_2}+\sqrt{s_3})^2$, 
$\lambda(s_1,s_2,s_3)=(s_1-s_1^R)(s_1-s_1^L)$, and
\begin{equation}
\phi_n(s)=\frac{G_1(s)P_n(s_1,s_2,s_3)\theta(s_1<s_1^-)}{(s_1-M_1^2)(s_1-s_1^L)^{n+1/2}}.
\end{equation}
For deriving this expression it is important that the functions $G_{1,2}$ have no 
singularities in the r.h.s. of the complex $s-$plane \cite{akms}. 
Along with the normal Landau--type contribution connected with the subprocess when all 
intermediate particles go on mass shell (the first term in (\ref{ff2})), 
the anomalous contribution 
(the second term in (\ref{ff2})) emerges at $q^2>0$ \cite{sr3}. 
The normal contribution dominates the form factor at small positive $q^2$ and vanishes at
the 'quark zero recoil point' $q^2=(m_2-m_1)^2$. The anomalous contribution is negligible at
small $q^2$ and grows as $q^2\to(m_2-m_1)^2$. 

We are now in a position to apply these results to the $B\to\pi,\rho$ decays, in which case 
$m_2=m_b$, $m_1=m_3=m_u$. To this end the quark model parameters ($m_u, m_b$ and the vertex functions)
should be specified. 

For a pseudoscalar meson built up of quarks with the masses $m_1$ and $m_2$, 
it is convenient to introduce the function $w$ related to the vertex function $G$ as follows
\begin{equation}
\label{5vertex}
G(s)=\frac{\pi}{\sqrt{2}}\frac{\sqrt{s^2-(m_1^2-m_2^2)^2}}
{\sqrt{s-(m_1-m_2)^2}}\frac{s-M^2}{s^{3/4}}w(k),\qquad
k=\frac{\lambda^{1/2}(s,m_1^2,m_2^2)}{2\sqrt{s}}
\end{equation}
The normalization condition (\ref{vertnorm}) for $G$ yields the following normalization 
condition for $w$
\begin{equation}
\int w^2(k)k^2dk=1.
\end{equation}
The function $w$ is the ground--state $S$--wave radial wave function of a 
pseudoscalar and vector meson for which a simple exponential form is usually chosen
\begin{equation}
\label{exp}
w(k)=\exp(-k^2/2\beta^2)
\end{equation}
We consider several sets of the quark model parameters 
used for the description of the meson spectra
and elastic form factors (Table \ref{table:parameters}). 

For analysing the results of calculations, introduce the functions 
$R_i(q^2)$ such that 
\begin{eqnarray}
\label{ffstech}
f_+(q^2)&=&R_+(q^2),   \nonumber \\
V(q^2)&=&(M_1+M_2)g(q^2)=(1+r)R_V(q^2),   \nonumber \\
A_1(q^2)&=&\frac1{(M_1+M_2)}f(q^2)=\frac{1+r^2-y}{1+r}R_1(q^2),  \nonumber \\
A_2(q^2)&=&-(M_1+M_2)a_+(q^2)=(1+r)\frac{1-r^2-y}{(1+r)^2-y}R_2(q^2),  \nonumber 
\end{eqnarray}
where $r=M_2/M_1$, $y=q^2/M_1^2$. 
As found by Stech \cite{stech}, $R_i(q^2)$ should be equal up to the corrections of the order 
$O\left(2m_{3}M_1/(M_1^2+M_2^2-q^2)\right)$. For the decay $B\to\pi$ the corrections 
about 10\% are expected even at $q^2=0$. 

Table \ref{table:fits} presents the parameters of the best fits to the calculated form factors 
in the form 
$$
R_i=\frac{R_i(0)}{(1-q^2/M_i^2)^{n_i}}.
$$ 
These fits approximate the form factors with better than 0.5\% accuracy in the relevant  
kinematic range and can be used for the calculation of the decay rates. 
In all the form factors, the anomalous contribution comes into the game only at 
$q^2\ge15\;$Gev${}^2$; below this point it is negligible. 
The decay rates are calculated with the form factors from Table \ref{table:fits}
via the formulas from Ref.\cite{gs}.   
Table \ref{table:results} summarises the results on the form factors and decay rates. 

The following conclusions can be drawn:

(i) Relativistic effects are important in the $B\to\pi,\rho$
decays, as our results considerably differ from those of the
models with nonrelativistic treating quark spins with the same 
parameters (Sets 1 and 2).

(ii) The functions $R_i$ are equal with an expected 10-20\%
accuracy as found by Stech for any set of the parameters. However, the magnitude 
and $q^2$-dependence of the functions $R$ strongly depend on these parameters (see Fig.2).
The 'quark model average values' for the decay rates can be determined as
\begin{eqnarray}
&\Gamma(\bar B^0\to\pi^+e\bar\nu)=(7\pm2)\times10^{12}|V_{ub}|^2\; s^{-1}, \nonumber \\ 
&\Gamma(\bar B^0\to\rho^+e\bar\nu)/\Gamma(\bar B^0\to\pi^+e\bar\nu)=1.45\pm0.1, \nonumber \\  
&\Gamma_L/\Gamma_T=0.7\pm0.08.
\end{eqnarray}
These values are not far from the predictions of other models (Table \ref{table:b2pirates}),
except for the results by Narison (\cite{sr4}). This is mainly due to
a very specific decreasing $q^2$--behavior of $A_1$ in \cite{sr4}. 
\footnote{The decay rates are rather sensitive to the details of the $q^2$--behaviour of $A_1$. 
In the light--cone approach the form factor $A_1$ has a specific feature: 
unlike other form factors it cannot be determined uniquely.   
Using the dispersion language, the form factor $A_1$ requires subtractions which cannot 
be fixed uniquely. This is a general situation for any amplitude with a vector particle 
in the initial or final state. Such an amplitude is transverse with respect to the 
momentum of the vector particle, and this yields a necessity of subtractions in some of 
the form factors. However, different definitions of the subtraction procedure 
lead to rather close results.}

The ratio of the decay rates have rather good accuracy, while their absolute values 
are less definite. Perhaps, the way out of this situation lies in a more detailed 
consideration of the constituent quark transition form factor. We use an approximation 
$f_{21}\equiv 1$. However, there are arguments in favour of a nontrivial $q^2$--dependence
of $f_{21}$. Actually, within the constituent quark picture, the constituent quark form
factor
arises from the two sources. First, it appears as a bare form factor which describes 
the constituent quark amplitude of the weak current defined through current quarks. 
One can assume this bare form factor to be close to unity at relevant $q^2$. Secondly, this
bare form factor is renormalised by the constituent quark rescatterings (final state
interactions) \cite{amn}. These very interactions yield meson formation and are not 
small at least in the region of $q^2$ near the meson mass. So the relation
$f_{21}=const$ cannot be valid for all $q^2$. 
Whether it is strongly violated or not at the $q^2$ of interest is not clear yet. 
Setting $f_{21}=1$ in (\ref{ff2}) means that we take into account the effects of the
constituent quark binding in the initial and final meson channels, and neglect the quark
rescatterings in the $q^2$--channel. The constituent quark form factor should depend on the
parameters of the model. One can expect that 
once the constituent quark form factor is taken into account thoroughly, the results on the meson
transition form factors and decay rates will be weakly dependent on the quark model parameters.  
In the ratio of the decay rates the constituent quark form factors cancel, 
and these results seem to be more reliable. 

I am grateful to V.~Anisovich, S.~Godfrey, I.~Grach, I.~Narodetskii, 
and K.~Ter--Martirosyan for valuable discussions and to V.~Nikonov for help 
in numerical calculations. The work was supported by the Russian Foundation for
Basic Research under grant 95--02--04808a.

\newpage
\begin{table}[1]
\caption{\label{table:b2piffs}
The form factors of the decays $B\to \pi,\rho$ at $q^2=0$. 
The labels QM, SR, and LAT stand for Quark Model, Sum Rules, and Lattice, respectively. }
\centering
\begin{tabular}{|c|c|c|c|c|c|c|c|}
\hline
\multicolumn{2}{|c|} {Ref.}& $f_+(0)$   & $V(0)$ & $A_1(0)$ & $A_2(0)$ & $V(0)/ A_1(0)$ & $A_2(0)/A_1(0)$ \\
 \hline\hline
 QM      &WSB \cite{wsb}  &  0.33  &  0.33   &   0.28  &  0.28   &  1.2  &    1.0   \\
         &ISGW \cite{isgw}&  0.09  &  0.27   &   0.05  &  0.02   &  5.4  &    0.4   \\
\hline
SR       & BBD \cite{sr1}    & 0.24$\pm$ 0.025 &           &            &             &           &         \\
         & Ball \cite{sr2}    & 0.26$\pm$ 0.02  &  0.6$\pm$0.2    &  0.5$\pm$0.1    &  0.4$\pm$0.2     &          &        \\
         & Narison \cite{sr4}    & 0.23$\pm$ 0.02  &  0.45$\pm$0.05  &  0.38$\pm$0.04  &  0.45$\pm$0.05   &
         1.11$\pm$ 0.01     &  1.11$\pm$ 0.01       \\         
\hline
Lat      & APE \cite{lat1}a & 0.29$\pm$ 0.06  & 0.45$\pm$0.22 & 0.29$\pm$0.16 & 0.24$\pm$ 0.56 & 2.0$\pm$0.9 & 0.8$\pm$1.5 \\
         & APE \cite{lat1}b & 0.35$\pm$ 0.08  & 0.53$\pm$0.31 & 0.24$\pm$0.12 & 0.27$\pm$ 0.80 & 2.6$\pm$1.9 & 1.0$\pm$3.1 \\

         & ELC \cite{lat2}a & 0.26$\pm$ 0.16  & 0.34$\pm$0.10 & 0.25$\pm$0.06 & 0.38$\pm$ 0.22 & 1.4$\pm$0.2 & 1.5$\pm$0.7 \\
         & ELC \cite{lat2}b & 0.30$\pm$ 0.19  & 0.37$\pm$0.11 & 0.22$\pm$0.05 & 0.49$\pm$ 0.26 & 1.6$\pm$0.3 & 2.3$\pm$0.9 \\
\hline\hline
\end{tabular}
\end{table}

\begin{table}[2]
\caption{\label{table:b2pirates}
Decay rates of the transitions $\bar B^0\to(\pi^+,\rho^+)e^-\bar\nu$ 
in units $|V_{ub}|^2\times 10^{12}\,s^{-1}$.}
\centering
\begin{tabular}{|c|c|c|c|c|}
\hline
Ref. &$\Gamma(B\to\pi)$ &$\Gamma(B\to\rho)$& $\Gamma_\rho/\Gamma_\pi$ & $\Gamma_L/\Gamma_T$ \\
 \hline\hline
WSB \cite{wsb}  &  7.43          &     26.1      &      3.5     &     1.34        \\
ISGW \cite{isgw}&  2.1           &     8.3       &      3.95     &    0.75        \\
\hline
Ball \cite{sr2}    & 5.1$\pm$ 1.1  &  12$\pm$4 &         &  0.06$\pm$0.02     \\
Narison \cite{sr4} & 3.6$\pm$ 0.6  &         & 0.9$\pm$0.2&  0.2$\pm$0.01      \\   
APE \cite{lat1}    & 8$\pm$4       &         &           &        \\
\hline\hline
\end{tabular}
\end{table}

\begin{table}[3]
\caption{\label{table:parameters}
Parameters of the quark model.}
\centering
\begin{tabular}{|c||c|c||c|c|}
\hline
Parameter Set & $m_u$ & $m_b$ & 
$\beta_{\pi}$ & $\beta_{B}$  \\
\hline
\hline
Set 1 \cite{wsb}     & 0.35  & 4.9  &  0.4   & 0.4  \\
Set 2 \cite{isgw}    & 0.33  & 5.12 &  0.31  & 0.41 \\
Set 3 \cite{jaus}    & 0.29  & 4.9  &  0.29  & 0.386\\
Set 4 \cite{schlumpf}& 0.25  & 4.7  &  0.32  & 0.55 \\
%Set 5 \cite{card}    & 0.22  & 4.977&  0.45* & 0.7* \\
\hline
\hline
\end{tabular}
%\tablenotes{* the variational solution of the GI model \cite{gi} found in 
%\cite{card} is used here}
\end{table}

\begin{table}[4]
\caption{\label{table:fits}
Parameters of the best fits to the calculated form factors.}
\centering
\begin{tabular}{|c||c|c|c||c|c|c||c|c|c||c|c|c|}
\hline
Set & $R_+(0)$ & $M_+$ & $n_+$ & $R_V(0)$ & $M_V$ & $n_V$ &
      $R_1(0)$ & $M_1$ & $n_1$ & $R_2(0)$ & $M_2$ & $n_2$ \\
\hline\hline
Set 1 &  0.29 & 6.29 & 2.35 & 0.30  & 6.28 & 2.36  
      &  0.27 & 7.07 & 2.65 & 0.25  & 6.13 & 2.17 \\
Set 2 &  0.20 & 6.22 & 2.45 & 0.20  & 6.22 & 2.46  
      &  0.20 & 6.78 & 2.65 & 0.19  & 6.00 & 2.34 \\
Set 3 &  0.21 & 5.90 & 2.33 & 0.21  & 5.90 & 2.35  
      &  0.21 & 6.50 & 2.70 & 0.20  & 5.90 & 2.45 \\
Set 4 &  0.26 & 5.44 & 1.72 & 0.29  & 5.46 & 1.73 
      &  0.29 & 5.68 & 1.67 & 0.28  & 5.36 & 1.67 \\ 
%Set 5 &  0.36 & 5.76 & 1.31 & 0.39  & 5.73 & 1.34  
%      &  0.35 & 5.61 & 1.10 & 0.31  & 4.82 & 0.88 \\ 
\hline
\hline
\end{tabular}
\end{table}

\begin{table}[5]
\caption{\label{table:results}
Form factors of the $\bar B^0\to(\pi^+,\rho^+)e^-\bar\nu$ decays 
at $q^2=0$ and the calculated decay rates 
in units $|V_{ub}|^2\times 10^{12}\;s^{-1}$.}
\centering
\begin{tabular}{|c||c|c|c|c|||c|c|c|c|}
\hline
Set & $f_+(0)$ & $V(0)$ & $A_1(0)$ & $A_2(0)$ & 
%$A_1/V$ & $A_2/V$ &
$\Gamma(B\to\pi)$ & $\Gamma(B\to\rho)$ &$\Gamma_\rho/\Gamma_\pi$ &$\Gamma_L/\Gamma_T$\\
\hline\hline
Set 1 &  0.29 & 0.34 & 0.24 & 0.21  & 9.2 & 12.6 & 1.38 & 0.77 \\
Set 2 &  0.20 & 0.22 & 0.17 & 0.16  & 5.0 & 7.25 & 1.45 & 0.73 \\
Set 3 &  0.20 & 0.24 & 0.19 & 0.17  & 6.1 & 8.8  & 1.44 & 0.62 \\
Set 4 &  0.26 & 0.33 & 0.26 & 0.24  & 8.9 & 13.8 & 1.55 & 0.67 \\
%Set 5 &  0.36 & 0.44 & 0.32 & 0.27  & 9.0 & 16.2 & 1.8  & 1.07 \\
\hline
\hline
\end{tabular}
\end{table}

\begin{figure}[1]
\begin{center}  
\mbox{\epsfig{file=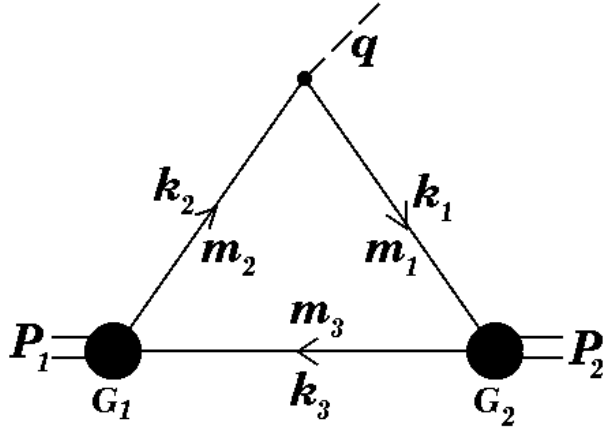,height=6.cm}  }
\end{center}
\caption{One-loop graph for a meson decay.\label{fig:1}}
\end{figure}

\begin{figure}[2]
\begin{center}  
\mbox{\epsfig{file=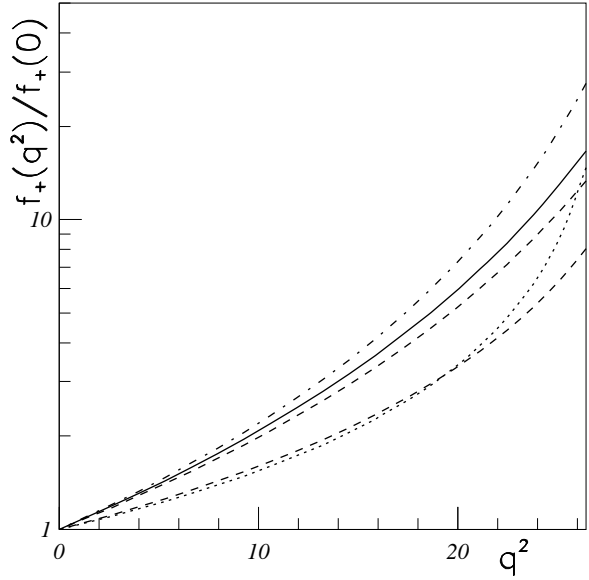,height=10.cm}  }
\end{center}
\caption{The $q^2$--dependence of the form factor $f_+$ for the decay 
$b\to\pi$ at various quark model parameters. 
Set 1 -- upper dashed line, set 2 -- lower solid line, 
set 3 -- dash--dotted line, set 4 -- lower dashed line, 
%set 5 -- upper solid line, 
the monopole formula with $M_{pole}=M_{B^*}=5.34\,$GeV -- dotted line. 
\label{fig:2}}
\end{figure}

\end{document}